\documentclass{emulateapj}
\usepackage{graphics}
\newcommand{\beq}{\begin{equation}}
\newcommand{\eeq}{\end{equation}}
\newcommand{\bea}{\begin{eqnarray}}
\newcommand{\eea}{\end{eqnarray}}
\newenvironment{inlinefigure}{
\def\@captype{figure}
\noindent\begin{minipage}{0.95\linewidth}\begin{center}}
{\end{center}\end{minipage}\smallskip}
\begin{document}

\title{The need for a second black hole at the Galactic center}
\author{Brad M. S. Hansen\altaffilmark{1} \& Milo\v s Milosavljevi\'c\altaffilmark{2}}
\altaffiltext{1}{Department of Astronomy, University of California Los Angeles, Los Angeles, CA 90095, hansen@astro.ucla.edu}
\altaffiltext{2}{Theoretical Astrophysics, California Institute of Technology, Pasadena, CA, 91125, milos@tapir.caltech.edu}
\righthead{SECOND BLACK HOLE AT THE GALACTIC CENTER}
\lefthead{HANSEN \& MILOSAVLJEVI\'C}
\begin{abstract}
Deep infra-red observations
and long-term monitoring programs
 have provided dynamical evidence for a supermassive
black hole of mass $3 \times 10^6  M_{\odot}$ 
associated with the
 radio source Sagitarrius A$^*$ at the center of our Galaxy.
 The brightest stars orbiting within 0.1 parsecs of 
the black hole appear to be young, massive main sequence stars,
in spite of an environment near the black hole that is hostile to star formation.
We discuss mechanisms by
which stars born outside the central parsec can sink towards 
the black hole and
conclude that the drag coming from
plausible stellar populations does not operate on the short timescales required by
the stellar ages. We propose that these stars
were dragged in by a second black hole of mass $\sim 10^{3-4} M_{\odot}$,
which would be classified as
an intermediate-mass black hole. We discuss the implications for the stellar populations
and the kinematics
in the Galactic center.
 Finally we note that continued astrometric monitoring of the central
radio source offers the prospect for
a direct detection of such objects.

\end{abstract}

\section{Introduction}

Near-infrared imaging observations with speckle and adaptive optics techniques now allow the study of
the central parsec of our Galaxy at an angular resolution of 0.05 arcseconds. Monitoring over the
course of the last decade has provided the proper motions for many of the stars orbiting in the central
parsec \citep{Eckart:97,Ghez:98}. 
Measurement of significant deviations from linear motion has yielded orbital solutions for several of the stars
closest ($<0.016$~pc) to the supermassive black hole (SBH) at the Galactic center\citep{Ghez:00,Eckart:02,Schoedel:02,Ghez:03a,Ghez:03b}.
While 
 the older, fainter stars exhibit an apparently isotropic distribution of
stellar velocities, there exist separate
populations of young stars, 
the motion of which appears to be dynamically unrelaxed.
At distances $\sim 0.1$ parsecs from the black hole lie a group of He emission-line stars (the IRS~16 cluster) belonging to a 
tangentially-anisotropic orbital family \citep{Genzel:00}. 
These stars, 
with K band magnitudes $\sim$9--14, 
are believed to be evolved, supergiant or Wolf-Rayet stars, with masses estimated at $30-100 \,  M_{\odot}$
after correcting for the distance and the extinction 
towards the Galactic center \citep{Ghez:03a}. 
As such, these stars can only be $\sim 1-10$ Myr
old.  Closer to Sgr~A$^*$ there is a coeval but apparently kinematically distinct population
of stars \citep{Genzel:97,Gezari:02,Ghez:03a,Ghez:03b,Genzel:03}.
 This group is claimed to be radially anisotropic and of an earlier, O/B spectral type. 
 It is the provenance of these two groups of stars and the origin of their peculiar kinematics that we wish to address.

\section{Previous Work}

In situ 
formation of young massive stars is unlikely inside the central parsec as the tidal forces 
 render it difficult for a sinking 
molecular cloud to survive
long enough to form stars close to the SBH. 
Clouds that are sufficiently dense to resist
tidal shear, $n\gtrsim 10^8$ cm$^{-3}$, are Jeans-unstable and fragment before they 
sink \citep{Vollmer:01}.
 An alternative suggestion \citep{Levin:03} is that the stars formed in an
extended,
self-gravitating 
gaseous disk that had in the past 
existed close to the SBH.  However
 there is no direct evidence to
corroborate this hypothesis. To explain the non-coplanar orbits of the 
central cluster, \citet{Levin:03} propose that the orbits of the closest
stars have been affected by Lens-Thirring precession due to an SBH 
spin that is not aligned with the disk axis.

If the young stars were not born in their current positions, they must
have migrated inward from
larger Galactocentric radii.
Massive star clusters
\citep{Gerhard:01,PZ:03,McMillan:03}
can sink via dynamical
friction within $\rm Myr$s but are tidally disrupted at a distance
$> 1 \, \rm pc$ from the
SBH where the specific gravitational binding energy is still $\sim 100$ times
smaller than that of the most bound He-line star SO-2.  Clusters impinging on the Galactic center on nearly-radial orbits are dispersed at their first passage near the SBH and result in a population of plunging, low-binding energy orbits unlike those of either the He line stars or of the central cluster.

Stars that form in the molecular clouds of the circum-nuclear disk (CND) at $\sim 1-2$ pc, as well as those that are deposited by the clusters disrupted at $\sim 1$ pc, diffuse toward orbits of larger binding energy 
on the local relaxation time scale:
 \bea
 T_{\rm rel} &\sim& 3\times 10^{9} {\rm yr} \, 
\left(\sigma \over 100~{\rm km~s}^{-1}\right)^3
\left(M_* \over 3M_\odot\right)^{-1}\nonumber\\
& &\times
\left(\rho \over 2\times10^5 M_\odot{\rm pc}^{-3}\right)^{-1}
\left(\frac{\ln\Lambda}{10}\right)^{-1} ,
\eea
where $\sigma$ is the local linear stellar velocity dispersion, $M_*$ is the average stellar mass, $\rho$ is the local stellar
density, and $\ln(\Lambda)$ is the Coulomb logarithm.
Not only does this exceed the
age of the observed stars, but relaxation via star-star
scattering would have also erased the observed kinematic peculiarities.

The fundamental reason for the long relaxation time is that the stellar
velocity dispersion $\sigma\propto R^{-1/2}$ diverges as the Galactocentric
radius $R$ decreases.
 Large velocities weaken 
the effect of gravitational focusing in stellar
scattering and decrease the average amount of energy transferred
in a single encounter. This limitation applies equally to all stellar mass
objects.
Only objects significantly more massive than a star can sink on the timescales required. Star clusters, however, are not dense
enough to survive intact in the strong tidal field \citep{PZ:03}. 
Thus, the only astrophysical
entity both massive and dense enough to satisfy the requirements is an intermediate-mass black hole (IBH).  
The orbital decay of an IBH is driven by dynamical friction on
a shorter timescale $T_{\rm df} \sim (M_*/M_{\rm IBH}) T_{\rm rel}\sim
1-10$~Myr for an $10^{3-4} M_{\odot}$ IBH. Furthermore, it has
been argued that such black holes form generically 
in dense, young stellar clusters
as a result of the segregation of massive 
stars to the cluster center \citep{Spitzer:69} followed by 
the runaway merging in stellar collisions \citep{PZ:02,Rasio:03}.

\section{Stellar Orbital Migration}

 A small fraction of the parent stellar cluster may remain 
bound to the IBH as it sinks in the Galactic potential, namely those
stars originally located within the dynamical radius of influence of the
IBH.
Tidal stripping and ejection due to strong stellar encounters gradually remove stars
from the cluster. In the absence of internal dynamical evolution, cluster stars
orbiting at distance $r$ from the IBH
will be lost when they slip over the cluster Roche limit 
 $r = (M_{\rm IBH}/M_{\rm SBH})^{1/3} R$, where $R$ is the distance to the SBH.

However, stars are also ejected by scattering off other stars in the cluster.
Using equation [37] of \citet{Lin:80}, the ejection timescale from a power-law cusp \citep{baw76} dominated
by the IBH is
\bea
 T_{\rm ej} &\sim& 1.4\times 10^6 {\rm yr} \, 
\left(\sigma_{\rm cl} \over 10~{\rm km~s}^{-1}\right)^3
\nonumber\\
& &\times
\left(M_* \over 10M_\odot\right)^{-1}
\left(\rho_{\rm cl} \over 10^5 M_\odot{\rm pc}^{-3}\right)^{-1}, 
\eea
where $\sigma_{\rm cl}$, $\rho_{\rm cl} $ are the velocity dispersion
and stellar mass density of the parent cluster at the original
dynamical radius of influence of the IBH.

The stars lost from the cluster are deposited over a range of radii.
If
stars are removed by tidal stripping alone, the profile of deposited
stars reflects the original profile of the cluster, $\rho(R) \sim R^{-\gamma}$,
where the Bahcall-Wolf value is $\gamma=7/4$. The stars most tightly bound
to the IBH will be lost at a distance from the SBH of
\bea
 R & \sim & 0.2 {\rm pc} \left( M_{\rm IBH} \over 10^3M_\odot\right)^{-7/3}
\left(M_* \over 10M_\odot\right)
\nonumber\\
& &\times
\left(\rho_{\rm cl} \over 10^5 M_\odot{\rm pc}^{-3}\right)^{-1}
\left(\sigma_{\rm cl} \over 10~{\rm km~s}^{-1}\right)^{4}
\eea

Therefore to deposit the most bound star SO-2 at $R\sim 0.01$ pc, 
$M_{\rm IBH}\sim 4\times 10^3 M_\odot$ is required.  
If the IBH orbital decay is eccentric and self-similar \citep{Valtaoja:89}, 
the profile may not be smooth; the young stars 
will be deposited in batches at a discrete set of locations corresponding to successive
pericenter passages of the IBH.

Even stars not directly bound to the IBH may 
be transported if their orbits come close that of the IBH. 
The dynamics of the interaction between stars and the SBH-IBH binary is very 
similar to the interaction of comets with the Sun-Jupiter binary. In the case
of the Jupiter family comets, they are observed in the inner solar system
after being scattered by Jupiter from orbits with much larger semi-major
axes \citep{Quinn:90}. In a similar fashion,
successive weak encounters with the IBH cause a random walk in the
orbital parameters, and this can nudge the peribothra of some field stars inward.
One difference is that the IBH is gradually moving inwards, in a fashion similar
to planetary migration, for which planetesimal/comet 
scattering may also be a contributing
factor \citep{Murray:98} in some extrasolar systems.
For this mechanism to be significant, the characteristic time 
$\mu^{-1/3} T_{\rm orb}$ between close
encounters with the IBH must be shorter than the time $\mu^{1/3} T_{\rm df}$ 
in which the IBH migrates a
distance equivalent to the radius of its own sphere of influence, where $T_{\rm orb}$ is the stellar orbital period
and $\mu=M_{\rm IBH}/M_{\rm SBH}$. 
This holds as long as
\beq
R<1 \, {\rm pc} \left( M_{\rm IBH} \over 10^3 M_{\odot} \right)^{-1/3}  
%\left( M_{\rm SBH} \over 3\times 10^6 M_{\odot} \right)^{4/3}
\left( \rho (1\, {\rm pc}) \over 2\times10^5 M_{\odot} {\rm pc}^{-3} \right)^{-1} ,
\eeq
where we have assumed the stellar density $\rho \propto 1/R^2$ on these scales.
Thus, the orbital decay of the IBH 
continues to push some stars inward even after the original cluster has 
been
tidally disrupted. 
The efficiency of these processes, however, is low, 
because most scattered stars are ultimately ejected---producing 
the Oort cloud in the case of comets 
\citep{Fernandez:81,Duncan:87}---and a large population 
is needed at the outset. On the other hand, 
for an IBH born of a $10^5 M_{\odot}$ cluster, the efficiency of the
process need only be $10^{-3}$ to explain the handful of IRS~16 stars.  

Dynamical friction drags the IBH toward the SBH until it reaches a critical separation where the binding energy
in stars with peribothra smaller than the SBH-IBH separation is the same as that of the IBH itself.
Using the most recent determination of central stellar cusp mass by \citet{Genzel:03},
$M_{\rm cusp} \sim 1.3 \times 10^4 M_{\odot} (R/1^{''})^{1.63}$, we infer this
critical radius to be $R \sim 0.2^{''} \sim 0.008 (M_{\rm IBH}/10^3 M_{\odot})^{0.61}$ parsecs.
 It could be smaller if there is a significant
dark stellar population interior to this orbit.
Subsequent orbital decay 
proceeds at a decreased rate 
contingent on how efficiently star-star
scattering and other forms of orbital diffusion feed stars into the emptied region (usually called the ``loss cone''), 
thereby providing new material for slingshot ejection
by the binary 
\citep{Begelman:80,Quinlan:96,Milosavljevic:03}.   
The binary shrinks by an octave for every $M_{\rm IBH}$-worth of stars that are removed.
Stars are usually ejected by the cumulative effect of several encounters, with their orbital parameters
undergoing a random walk. 
If the separation decreases to $10\, {\rm AU}\,(M_{\rm IBH}/10^3M_\odot)$, the emission of
 gravitational radiation dominates and will expedite the coalescence of the black holes within a million years.
%\newpage
\section{The distribution of stars in the inner parsec}
\label{inner}

If the He-line stars are dynamically associated
with the IBH, then their orbits must either be similar to the orbit the IBH 
had at some, perhaps earlier, stage of its infall,
or must at least still cross the latter 
if they 
are in the process of being ejected by repeated scattering. 
To test this, we have examined the existing velocity
data (\citet{Genzel:00};A. Ghez,private communication)
for He-line stars with three measured velocity components. 
Combined with the projected position,
we have information on 5 of the 6 phase space coordinates 
necessary to compute the orbits, as well as complete solutions for a few close
to the SBH\citep{Ghez:03b} . 
To assess the importance of
the unknown component (location along the line of sight, $Z$) 
we explored the range from
$Z=0.5 \, R_{\perp}$ to $Z=1.5 \, R_{\perp}$, 
where $R_{\perp}$ is the projected distance to
Sgr~A$^*$. Figure \ref{figure1} shows the resulting orbits which are
plotted in a specific energy-angular momentum diagram.

\begin{inlinefigure}
\begin{center}
\resizebox{\textwidth}{!}{\includegraphics{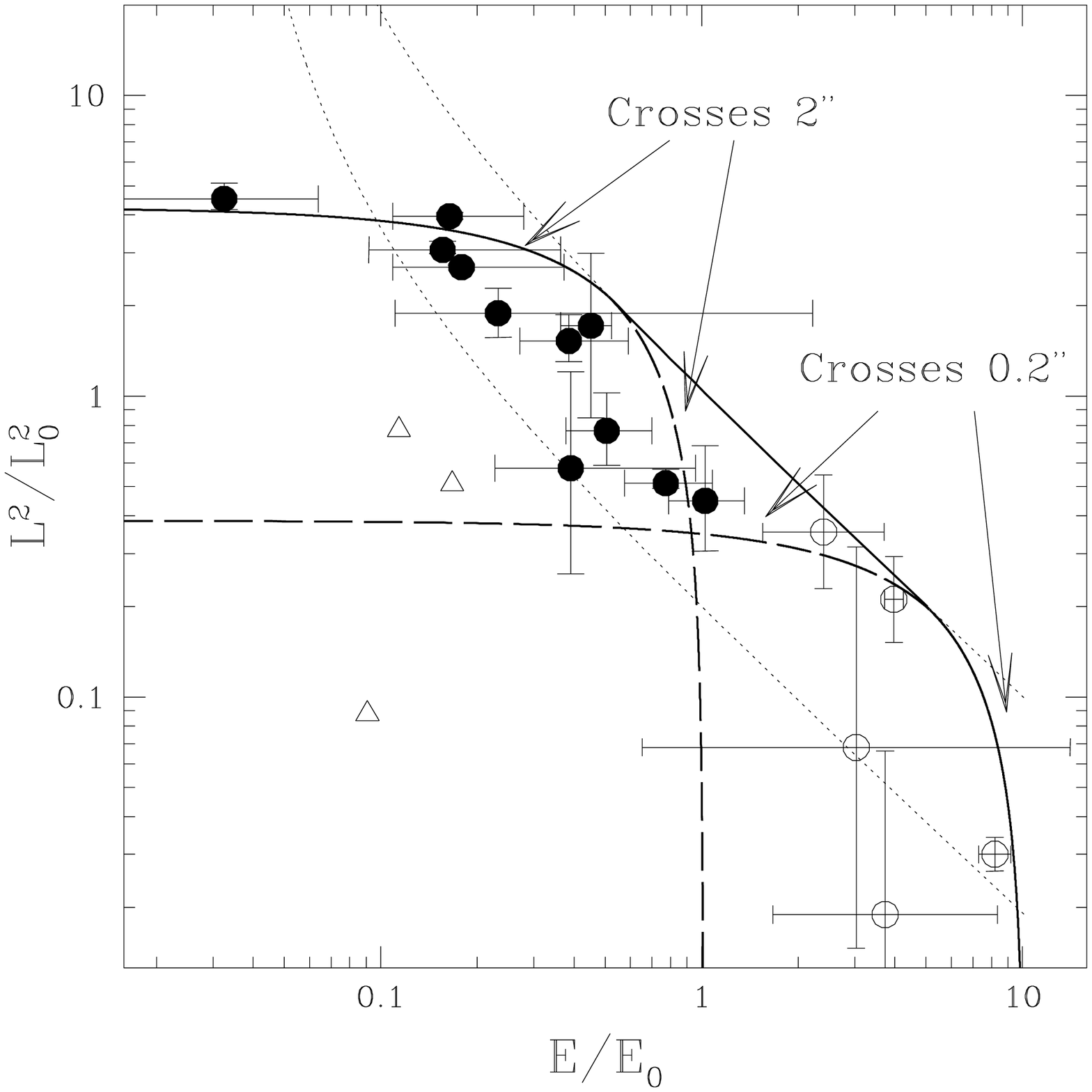}}
\end{center}
%\plotone{fig1.ps}
\figcaption{The filled circles show the  He-line stars
associated with the IRS~16 group. The error bars indicate the effect of varying the unknown
$z$ between $z=0.5 r_{\perp}$ and $z=1.5 r_{\perp}$. The open circles
show the central cluster stars for which complete orbits can now be determined
\citep{Ghez:03a}
(thus these error bars are real). Triangles indicate approximate values for central
cluster stars for which complete orbits are not yet known. The solid line indicates the locus
of orbits which cross an eccentric orbit with semi-major axis $1''$ and eccentricity 0.82. 
An
IBH on this orbit can interact with all the stars shown.
 The dashed lines indicate the same
for circular orbits at $2''$ and $0.2''$ instead. The dotted lines indicate
the locus of circular orbits (rightmost) and orbits with eccentricity 0.9. $E_0$ and $L_0$
are calculated assuming an SBH mass
of $3 \times 10^6 M_{\odot}$ and a semi-major axis of $1''=0.04$~parsecs.
\label{figure1}}
\end{inlinefigure}

The IRS~16 stars all cross a circular orbit located 2$^{''}$ from Sgr~A$^*$.
As such they are consistent with being the remnant of a disrupted IBH host cluster or, alternatively,
are currently undergoing the scattering-induced diffusion and the eventual ejection if the
IBH is located at this radius. The more tightly bound
stars, the Sgr~A$^*$ cluster, do not cross this orbit. However, it is possible for an eccentric IBH orbit to interact
with both the IRS~16 stars and the Sgr~A$^*$ cluster. The solid curve in Figure \ref{figure1} delineates that region of parameter
space (that part below the curve) in which stars interact with an orbit of semi-major axis of $1^{''}$(0.04~pc) and
 eccentricity 0.82, which covers all the early-type stars.

It is also possible that the IBH orbit is indeed not very eccentric, and that the central cluster stars once did cross the IBH orbit, but have since then
 received a significant perturbation near peribothron that resulted in a new apbothron within the orbit
of the IBH. The star is then trapped until further IBH inspiral brings the orbits into contact again.
The claimed radial anisotropy of these stars \citep{Schoedel:03} is consistent with such a scenario.
The perturbation could have come from a
grazing stellar collision or from the dynamical disruption of a binary system \citep{Gould:03}.
  For the perturbation to be strong enough to trap the star in a closer-in orbit,
 the collision must be grazing because the velocities in this region are comparable to the
escape velocities from the stars themselves, i.e., the dynamical evolution is dominated by
physical collisions rather than two-body relaxation. 

We find that
the spectral differences between the central cluster stars and the IRS~16 stars lend support to this picture. Although
both are young and massive, the former appear to be main sequence
stars of roughly O or B type, while the latter are extended, evolved stars in the supergiant or the Wolf-Rayet
stage. The more compact nature of the central cluster stars may indicate that the outer envelopes have been
stripped in grazing collisions.
Note that this is different from the alternative proposal (e.g., \citealt{Genzel:03}) in which the observed stars are
assembled from lower mass objects in stellar mergers. 
The normal rotation rate of SO-2 \citep{Ghez:03a} and the potential radial
anisotropy of the orbits 
are both consistent with a limited role of collisional interactions.

If an IBH is present at the Galactic center, there is also the
possibility of directly observing high velocity stars 
that have been ejected from the region.  This would especially be true if the SBH-IBH binary has hardened past the stalling radius and is continuing to evolve slowly via the gravitational slightshot of stars that diffuse into the loss cone.  The velocities of the ejected stars will reflect the specific binding energy of the IBH at the time of ejection. 

Direct constraint on the IBH hypothesis may be provided by the
 non-inertial motion of Sgr~A$^*$. 
 The presence of an IBH at the Galactic center will be
revealed in the apparent proper motion of the radio image of Sgr A$^*$ 
relative to the mean $\sim 6$ mas yr$^{-1}$ solar drift 
\citep{Backer:99,Reid:99}.  Assuming a circular orbit of the IBH, the
amplitude of the gravitational reflex of SBH relative to the barycenter of the system as a whole,
$\Psi_{\rm ref} \sim  M_{\rm IBH}/M_{\rm SBH}\, \Psi$, where $\Psi$ is the angular separation between the black
holes, must be larger than the cumulative 
positional resolution $\Delta \Psi<1$~mas of the
radio telescope to achieve detection.  In turn, the total projected distance
traversed by the SBH, equal to $\Psi_{\rm ref} \sim  T_{\rm m} M_{\rm IBH} R_{\rm gc}^{-1} (G/R M_{\rm SBH})^{1/2}$ for
 a monitoring program of duration $T_{\rm m}$ and distance to the Galactic center $R_{\rm gc}\sim 8$~kpc, must also be larger than $\Delta \Psi$.
These constraints are summarized in Figure \ref{figure2}.

\section{Conclusions}

In this paper we have proposed a model that addresses the peculiar nature
of the young stars in the Galactic center. Most importantly, the presence
of an IBH can 
 deliver the massive young stars to
their observed location within the timescale required. In addition, our
scenario provides a way to link the two disparate groups of stars 
(the IRS~16 group and the Sgr~A$^*$ cluster) within a single
evolutionary scenario.
\begin{inlinefigure}
%\plotone{fig2.ps}
\begin{center}
\resizebox{\textwidth}{!}{\includegraphics{fig2.ps}}
\end{center}
\figcaption{
Assuming a circular orbit around an SBH of $3 \times 10^6 M_{\odot}$, we
can rule out an IBH with mass $M_{\rm IBH}$ 
and semi-major axis $a$ by measurement
of an astrometric wobble of the radio image of Sgr A$^*$.
The shaded regions show the detection thresholds
for astrometric resolutions of $0.01$, $0.1$ and 1~milliarcseconds,
respectively, assuming a monitoring period of 10 years. The dashed lines
indicate coalescence due to gravitational radiation in $10^6$ and
$10^7$~years, respectively.
\label{figure2}}
\end{inlinefigure}

In this scenario, once the IBH inspiral stalls, the stellar orbits continue
to evolve under the action of scattering during close encounters with the IBH.
Most stars will eventually be ejected, but some stars may be trapped near the
SBH if they receive a sufficiently large perturbation (such as due to a grazing
stellar collision). This may also serve to generate the observed spectral differences
between the extended and apparently undisturbed He-line stars, and the more
compact but apparently contemporaneous Sgr~A$^*$ cluster stars. Furthermore, 
the tangential anisotropy of the He line stars may reflect their still somewhat
limited dynamical evolution under scattering, whereas the claimed radial anisotropy
of the Sgr~A$^*$ cluster stars may reflect their having diffused to highly eccentric
orbits before being scattered and trapped. However, this radial anisotropy is still
uncertain because its determination is subject to severe selection effects. 

Some questions remain open: Although the mass density of luminous stars can
be measured, 
 the true density profile
of the central parsec is not well-known.
Our adoption of the observed density profile is conservative as
there are good
reasons to believe that $10M_\odot$ black holes will have segregated 
within few
relaxation times to
dominate the density of the central cluster 
\citep{Morris:93,meg:00}.
If such a dark cusp exists the central density could be as large as 
$10^9 M_\odot {\rm pc}^{-3}$ and thus the stalling radius of the IBH could be
much smaller than estimated above.
Alternatively, if the IBH infall is an episodically recurring phenomenon, then
the injection of a steady
stream of IBH into the central parsec would help to maintain 
an evacuated region in the stellar distribution.

Finally, IBH are implicated
in the formation of SBH in galaxies in a 
variety of ways \citep{Rees:84}. Indeed, it has been 
suggested \citep{Ebisuzaki:01b} that SBH form
from the collapse of a cluster of IBH. It is intriguing that to gradually accumulate the present SBH at the Galactic
center by this process requires IBH captures at the rate
\beq
\Gamma \sim 3 \times 10^{-7} {\rm yr}^{-1} \left( M_{\rm SBH} \over 3 \times 10^6 M_{\odot} \right) 
\left( M_{\rm IBH} \over 10^3 M_{\odot} \right)^{-1} .
\eeq
With this rate one would expect the most recent
entrant to be a few Myr old, which coincides with the time scale derived above from the stellar ages.
Such a process could have a significant influence on the history of galactic
nuclei \citep{Hughes:03}.

\acknowledgements 
The authors would like to thank Phil Armitage, Andrea Ghez, Mark Morris, Simon Portegies Zwart, Fred Rasio, Angelle
Tanner and Scott Tremaine for comments, and David Merritt and Sterl Phinney for valuable discussions.  MM thanks the Sherman Fairchild Foundation for support.


\begin{thebibliography}{99}

\bibitem[\protect\astroncite{{Backer} \& {Sramek}}{1999}]{Backer:99}
{Backer}, D.~C. \& {Sramek}, R.~A., 1999,
\newblock {\em \apj} { 524}, 805

\bibitem[\protect\astroncite{{Bahcall} \& {Wolf}}{1976}]{baw76}
{Bahcall}, J.~N. \& {Wolf}, R.~A., 1976,
\newblock {\em \apj} { 209}, 214

\bibitem[\protect\astroncite{{Begelman}, {Blandford} \& {Rees}}{1980}]{Begelman:80}
{Begelman}, M.~C., {Blandford}, R.~D., \& {Rees}, M.~J., 1980,
\newblock {\em \nat} { 287}, 307

\bibitem[\protect\astroncite{{Duncan}, {Quinn} \& Tremaine}{1987}]{Duncan:87}
{Duncan}, M., {Quinn}, T., \& {Tremaine}, S., 1987,
\newblock {\em \aj} { 94}, 1330

\bibitem[\protect\astroncite{{Ebisuzaki} {\rm et~al.\/}}{2001}]{Ebisuzaki:01b}
{Ebisuzaki}, T., et~al.~2001,
\newblock {\em \apjl} { 562}, L19

\bibitem[\protect\astroncite{{Eckart} et al. }{2002}]{Eckart:02}
{Eckart}, A., {Genzel}, R., {Ott}, T. \& Sch\"{o}del, R., 2002,
\newblock {\em \mnras} { 331}, 917

\bibitem[\protect\astroncite{{Eckart} \& Genzel }{1997}]{Eckart:97}
{Eckart}, A., \& {Genzel}, R., 1997
\newblock {\em \mnras} { 284}, 576

\bibitem[\protect\astroncite{{Fernandez} \& {Ip}}{1981}]{Fernandez:81}
{Fernandez}, J.~A. \& {Ip}, W.-H., 1981,
\newblock Icarus {47}, 470

\bibitem[\protect\astroncite{{Genzel} {\rm et~al.\/}}{1997}]{Genzel:97}
{Genzel}, R., {Eckart}, A., {Ott}, T., \& {Eisenhauer}, F., 1997,
\newblock {\em \mnras} { 291}, 219

\bibitem[\protect\astroncite{{Genzel} {\rm et~al.\/}}{2000}]{Genzel:00}
{Genzel}, R., {Pichon}, C., {Eckart}, A., {Gerhard}, O.~E., \& {Ott}, T., 2000,
\newblock {\em \mnras} { 317}, 348

\bibitem[\protect\astroncite{{Genzel} {\rm et~al.\/}}{2003}]{Genzel:03}
{Genzel}, R., et~al.~2003, astro-ph/0305423

\bibitem[\protect\astroncite{{Gerhard}}{2001}]{Gerhard:01}
{Gerhard}, O., 2001,
\newblock {\em \apjl} { 546}, L39

\bibitem[\protect\astroncite{{Gezari} {\rm et~al.\/}}{2002}]{Gezari:02}
{Gezari}, S., Ghez, A. M., Becklin, E. E., Larkin, J., McLean, I. S. \& Morris, M., 2002,
\newblock {\em \apj} { 576}, 790

\bibitem[\protect\astroncite{{Ghez} {\rm et~al.\/}}{2000}]{Ghez:00}
{Ghez}, A.~M., {Morris}, M., {Becklin}, E.~E., {Tanner}, A. \& {Kremenek}, T., 2000,
\newblock {\em \nat} { 407}, 349

\bibitem[\protect\astroncite{{Ghez} {\rm et~al.\/}}{2003a}]{Ghez:03b}
{Ghez}, A.~M., {Becklin}, E., {Duch{\^ e}ne}, G., 
{Hornstein}, S., {Morris}, M.,
  {Salim}, S., \& {Tanner}, A., 2003a, astro-ph/0303151

\bibitem[\protect\astroncite{{Ghez} {\rm et~al.\/}}{2003b}]{Ghez:03a}
{Ghez}, A.~M., et~al.~2003b,
\newblock {\em \apjl} { 586}, L127

\bibitem[\protect\astroncite{{Ghez} {\rm et~al.\/}}{1998}]{Ghez:98}
{Ghez}, A.~M., {Klein}, B.~L., {Morris}, M., \& {Becklin}, E.~E., 1998,
\newblock {\em \apj} { 509}, 678

\bibitem[\protect\astroncite{{Gould} \& {Quillen}}{2003}]{Gould:03}
{Gould}, A. \& {Quillen}, A.~C., 2003,
astro-ph/0302437

\bibitem[\protect\astroncite{{Hughes} \& {Blandford}}{2003}]{Hughes:03}
{Hughes}, S.~A. \& {Blandford}, R.~D., 2003,
\newblock {\em \apjl} { 585}, L101

\bibitem[\protect\astroncite{{Levin} \& {Beloborodov}}{2003}]{Levin:03}
{Levin}, Y. \& {Beloborodov}, A.~M., 2003,
astro-ph/0303436

\bibitem[\protect\astroncite{{Lin} \& {Tremaine}}{1980}]{Lin:80}
{Lin}, D.~N.~C. \& {Tremaine}, S., 1980,
\newblock {\em \apj} { 242}, 789

\bibitem[\protect\astroncite{{McMillan} \& {Portegies
  Zwart}}{2003}]{McMillan:03}
{McMillan}, S. \& {Portegies Zwart}, S.~F., 2003,
astro-ph/0304022

\bibitem[\protect\astroncite{{Milosavljevi\'c} \&
  {Merritt}}{2002}]{Milosavljevic:03}
{Milosavljevi\'c}, M. \& {Merritt}, D., 2003,
astro-ph/0212270

\bibitem[\protect\astroncite{{Miralda-Escud{\' e}} \& {Gould}}{2000}]{meg:00}
{Miralda-Escud{\' e}}, J. \& {Gould}, A., 2000,
\newblock {\em \apj} { 545}, 847

\bibitem[\protect\astroncite{{Morris}}{1993}]{Morris:93}
{Morris}, M., 1993,
\newblock {\em \apj} { 408}, 496

\bibitem[\protect\astroncite{{Murray} {\rm et~al.\/}}{1998}]{Murray:98}
{Murray}, N., {Hansen}, B., {Holman}, M., \& {Tremaine}, S., 1998,
\newblock {Science} { 279}, 69

\bibitem[\protect\astroncite{{Portegies Zwart}, {McMillan} \& {Gerhard}}{2003}]{PZ:03}
{Portegies Zwart}, S.~F., {McMillan}, S., \& {Gerhard}, O., 2003,
astro-ph/0303599

\bibitem[\protect\astroncite{{Portegies Zwart} \& {McMillan}}{2002}]{PZ:02}
{Portegies Zwart}, S.~F. \& {McMillan}, S.~L.~W., 2002,
\newblock {\em \apj} { 576}, 899

\bibitem[\protect\astroncite{{Quinlan}}{1996}]{Quinlan:96}
{Quinlan}, G.~D., 1996,
\newblock {NewA} { 1}, 35

\bibitem[\protect\astroncite{{Quinn}, {Tremaine} \& {Duncan}}{1990}]{Quinn:90}
{Quinn}, T., {Tremaine}, S., \& {Duncan}, M., 1990,
\newblock {\em \apj} { 355}, 667

\bibitem[\protect\astroncite{{Rasio}, {Freitag} \& {G{\" u}rkan}}{2003}]{Rasio:03}
{Rasio}, F.~A., {Freitag}, M., \& {G{\" u}rkan}, M.~A., 2003,
astro-ph/0304038

\bibitem[\protect\astroncite{{Rees}}{1984}]{Rees:84}
{Rees}, M.~J., 1984,
\newblock {\em \araa} { 22}, 471

\bibitem[\protect\astroncite{{Reid} {\rm et~al.\/}}{1999}]{Reid:99}
{Reid}, M.~J., {Readhead}, A.~C.~S., {Vermeulen}, R.~C., \& {Treuhaft}, R.~N.,
  1999,
\newblock {\em \apj} { 524}, 816

\bibitem[\protect\astroncite{{Sch{\" o}del} {\rm et~al.\/}}{2002}]{Schoedel:02}
{Sch{\" o}del}, R., et~al.~2002,
\newblock {\nat} { 419}, 694

\bibitem[\protect\astroncite{{Sch\"odel} {\rm et~al.\/}}{2003}]{Schoedel:03}
{Sch\"odel}, R., {Genzel}, R., {Ott}, T., \& {Eckart}, A., 2003,
astro-ph/0304197

\bibitem[\protect\astroncite{{Spitzer}}{1969}]{Spitzer:69}
{Spitzer}, L.~J., 1969,
\newblock {\em \apjl} { 158}, L139

\bibitem[\protect\astroncite{{Valtaoja}, {Valtonen} \& {Byrd}}{1989}]{Valtaoja:89}
{Valtaoja}, L., {Valtonen}, M.~J., \& {Byrd}, G.~G., 1989,
\newblock {\em \apj} { 343}, 47

\bibitem[\protect\astroncite{{Vollmer} \& {Duschl}}{2001}]{Vollmer:01}
{Vollmer}, B. \& {Duschl}, W.~J., 2001,
\newblock {\em \aap} { 377}, 1016

\end{thebibliography}
\end{document}